\documentclass[twocolumn,aps,prd,eqsecnum,showpacs]{revtex4}
\usepackage{graphicx}
\def\apjs{Astrophys.\ J.\ Suppl.\ }
\def\aa{Astron.\  Astrophys.\ }
\def\cqg{Class.\ Quantum Grav.\ }
\newcommand{\ddp}[2]{\frac{\partial #1}{\partial #2}}
\newcommand{\ddps}[2]{\frac{\partial^2 #1}{\partial #2 ^2}}
\begin{document}
\title
{Viscosity driven instability in rotating relativistic stars}
%
\author{Motoyuki Saijo}
\email{ms1@maths.soton.ac.uk}
%
\affiliation
{School of Mathematics, University of Southampton, 
Southampton SO17 1BJ, United Kingdom}
%
\author{Eric Gourgoulhon}
\email{eric.gourgoulhon@obspm.fr}
%
\affiliation
{Laboratoire de l'Univers et de ses Th\'eories, UMR 8102
du CNRS,
Observatoire de Paris, F-92195 Meudon Cedex, France}
%
\received{21 June 2006}
\revised{4 September 2006}
\accepted{8 September 2006}
%
\begin{abstract}
We investigate the viscosity driven instability in rotating
relativistic stars by means of an iterative approach.  We focus on
polytropic rotating equilibrium stars and impose an $m=2$ perturbation
in the lapse.  We vary both the stiffness of the equation of state and
the compactness of the star to study those effects on the value of the
threshold.  For a uniformly rotating star, the criterion $T/W$,
where $T$ is the rotational kinetic energy and $W$ is the
gravitational binding energy, mainly depends on the compactness of the
star and takes values around $0.13 \sim 0.16$, which differ slightly
from that of Newtonian incompressible stars ($\sim 0.14$).  For
differentially rotating stars, the critical value of $T/W$ is found to
span the range $0.17 - 0.25$.  This is significantly larger than the
uniformly rotating case with the same compactness of the star.
Finally we discuss a possibility of detecting gravitational waves from
viscosity driven instability with ground-based interferometers.
\end{abstract}
%
\maketitle
%
\section{Introduction}

Stars in nature are usually rotating and subject to nonaxisymmetric
rotational instabilities (see \citep{Sterg03}, \citep{Ander03},
\citep{AnderC06} or \citep{Villa06} for recent reviews).  An exact
treatment of these instabilities exists only for incompressible
equilibrium fluids in Newtonian gravity,
(e.g. \citep{Chandra69,Tassoul78,ST83}).  For these configurations,
global rotational instabilities arise from non-radial toroidal modes
$e^{im\varphi}$ ($m=\pm 1,\pm 2, \dots$) when $\beta \equiv T/W$
exceeds a certain critical value. Here $\varphi$ is the azimuthal
coordinate and $T$ and $W$ are the rotational kinetic and
gravitational binding energies.  In the following we will focus on the
$m=\pm 2$ bar-mode, since it is the fastest growing mode when the
rotation is sufficiently rapid.

There exist two different mechanisms and corresponding timescales for
bar-mode instabilities.  Uniformly rotating, incompressible stars in
Newtonian theory are {\em secularly} unstable to bar-mode formation
when $\beta \gtrsim \beta_{\rm sec} \simeq 0.14$.  This instability
can grow in the presence of some dissipative mechanism, like viscosity
or gravitational radiation, and the growth time is determined by the
dissipative timescale, which is usually much longer than the dynamical
timescale of the system.  By contrast, a {\em dynamical} instability
to bar-mode formation sets in when $\beta \gtrsim \beta_{\rm dyn}
\simeq 0.27$.  This instability is independent of any dissipative
mechanism, and the growth time is the hydrodynamic timescale of the
system.

There are two representative dissipative mechanisms that drive the
secular bar mode instability, viscosity and the gravitational
radiation, in the absence of thermal dissipation.  The viscosity
driven instability sets in when a mode has a zero-frequency in the
frame rotating with the star \citep{RS63}, and the first unstable mode
in terms of $m$ is the $m=2$ bar mode.  The quasi-static evolution of
the star due to viscosity driven instability, which varies the
circulation of the star, deforms the Maclaurin spheroid to the Jacobi
ellipsoid in Newtonian incompressible stars.  On the other hand, the
gravitational radiation induced instability \citep{Chandra70, FS78}
sets in when the backward going mode is dragged forward in the
inertial frame (see \citep{Sterg03,AnderC06,Villa06,Schutz87} for
reviews), and the mode in terms of $m$ are all unstable when it
exceeds a certain $m$.  The quasi-static evolution of the star due to
gravitational radiation induced instability, which varies the angular
momentum of the star, deforms the Maclaurin spheroid to the Dedekind
ellipsoid in Newtonian incompressible stars.

The viscosity driven instability, especially to determine the critical
value either in an incompressible star or in an ellipsoidal
equilibrium, has been studied in Newtonian gravity \citep{IM81}, in
post Newtonian gravity \citep{SZ98, GV02}, and in full general
relativity \citep{BFG96, BFG98, GG02} by an ellipsoidal approximation
(e.g. \citep{Chandra69}) or by an iterative evolution approximation
(e.g. \citep{BFG96}), and shows that the viscosity drives the
instability to higher rotation rates $\beta_{\rm sec} \gtrsim 0.14$ as
the configurations become more compact.  There is also a study of
Newtonian compressible stars to determine the critical value of
viscosity driven instability \citep{BFG96}: It is found that the star
becomes secularly unstable at $\beta_{\rm sec} \approx 0.135 \pm
0.02$, depending on the stiffness of the polytropic equation of
state.

The aim of the paper is twofold.  One is to investigate the
critical value of viscosity driven instability in the compressible
stars rotating uniformly.  The argument that
the viscosity driven instability plays a role to deform a star from a
Maclaurin spheroid to a Jacobi ellipsoid is only true in the absence
of internal energy, since the total energy has a chance to transfer it
to the internal energy without emission from the star
(e.g. \citep{Shapiro04}).  Here we assume that the cooling timescale
of the star is shorter than the thermal heating timescale so that the
thermal energy generated by viscosity is immediately radiated away.
Therefore the picture of the deformation process due to viscosity is
quite similar to the case of incompressible stars.  The stars are
usually considered as compressible bodies, and therefore it is worth
to take such effect into account whether there is a significant change
on the threshold.  In this respect the present work extends that of
Ref.~\citep{GG02} to the compressible case. 

The other aim of the present work is to investigate the effect of
differential rotation on the viscosity driven secular bar mode
instabilities.  For a strong viscosity or strong magnetic field
circumstances, the star maintains uniform rotation.  However, in
nature the star may rotate differentially as the Sun.  Stellar
collapses and mergers may also lead to differentially rotating stars
(e.g. \citep{Ott04}).  For the coalescence of binary irrotational
neutron stars \citep{SU00,SU02,STU03}, the presence of differential
rotation may temporarily stabilize the ``hypermassive'' remnant, which
constructs a differential rotation.  Therefore it is also worth to
take differential rotation into account to study viscosity driven
instabilities in rotating relativistic stars.

This paper is organized as follows.  In Sec.~\ref{sec:bequation} we
present the basic equations of our treatment of general relativity. We
discuss our numerical results in Sec.~\ref{sec:rigid} and in
Sec.~\ref{sec:diff}, focusing on the viscosity driven instability in
uniformly and differentially rotating stars.  In
Sec.~\ref{sec:Discussion} we briefly summarize our
findings. Throughout this paper, we use the geometrized units with
$G=c=1$ and adopt polar coordinates $(r,\theta,\varphi)$ with the
coordinate time $t$.  Note that Latin index takes
$(r,\theta,\varphi)$.

\section{Iterative evolution approach to determine the threshold of
  viscosity driven instability} 
\label{sec:bequation}
\subsection{Equilibrium configuration of rotating relativistic stars}
We briefly introduce our basic approach to construct rotating
relativistic stars and allow nonaxisymmetric deformation induced by
``viscosity''.   We introduce a nonaxisymmetric line element, when the
azimuthal component $\varphi$ is separable in the metric
\citep{Chandra83}, in spherical coordinates $(t, r, \theta, \varphi)$
with a quasi-isotropic gauge (e.g. \citep{BGSM93, GHLPBM99}) as
\begin{eqnarray}
ds^{2} &=& - N^{2} dt^{2} + A^{2} (dr - N^r dt)^2 + 
  r^2 A^2 (d\theta - N^{\theta} dt)^{2} 
\nonumber \\
&&
+ 
  r^{2} \sin^{2} \theta B^{2} (d\varphi - N^{\varphi} dt)^{2}
,
\end{eqnarray}
where $N$ is the lapse, $N^{r}$, $N^{\theta}$, $N^{\varphi}$
correspond to the shift, $A$ and $B$ are the spatial metric functions.
In the equilibrium state, we only take the azimuthal component of the
shift $N^{\varphi}$, the lapse, and two spatial metric functions into
account of the metric components.  Note that all of them are functions
of $r$ and $\theta$ only.  Once we impose the nonaxisymmetric
perturbation in the lapse, we take all the components of shift into 
account, and relax the dependence of the functions of ($r$, $\theta$)
to ($r$, $\theta$, $\varphi$) for lapse and shift \citep{BFG98}.  In
our present approach the nonaxisymmetric terms have been taken into
account at least to the $1/2$ post-Newtonian order.  Note that the
spatial metric functions keep the functional dependence of ($r$,
$\theta$) as those terms are considered as a higher post-Newtonian
order which only enters in the order of $\varepsilon_{\rm amp} \times
(M/R) \sim 10^{-6}$, where $\varepsilon_{\rm amp}$ is the amplitude of
the perturbation we imposed in the lapse, $M$ the gravitational mass,
$R$ the circumferential radius. It is therefore a good approximation
to drop the nonaxisymmetric contribution from the spatial part of the 
metric.  To summarize we compute the exact relativistic rotating
equilibrium star and perturb the geometrical quantities in lapse and
in shift, neglecting the azimuthal perturbation in the spatial
metric.

We also adopt the maximal slicing condition, for which the trace of the
extrinsic curvature $K_{ij}$ vanishes
\begin{equation}
K \equiv \gamma^{ij} K_{ij} = 0.
\end{equation}  
The gravitational field equations (Einstein equations) for the six
unknown functions $N$, $N^{r}$, $N^{\theta}$, $N^{\varphi}$,   $A$,
$B$ are written as \citep{BFG98}
\begin{eqnarray}
& & \bar{\Delta} \, \nu \; = \; 
   4 \pi A^2 [ E + 3 P + (E + P) U_i U^i ] + A^{2} K_{ij} K^{ij} 
\nonumber \\
&& \quad
   - \bar{\nabla}_i \nu \, \bar{\nabla}^i (\nu + \beta),
\label{e:Einstein1} \\
& & \bar{\Delta} \, N^{i} 
   + \frac{1}{3} \bar{\nabla}^i \bar{\nabla}_j N^j  \; = \;  
   - 16 \pi N A^2 (E + P) U^i 
\nonumber \\
&& \quad
   + N B^{-2} K^{ij} \bar{\nabla}_{j} (6 \beta - \nu) 
   \equiv J^i ,
\label{e:Einstein2} \\
& & \Delta_2 \, [ r \sin\theta (N B - 1)]  \; = \; 
    16 \pi r \sin\theta N A^2 B P,
\label{e:Einstein3} \\
& & \Delta_2 \, \zeta \; = \; 
  8\pi  A^2 \, [P + (E+P) U_i U^i] + \frac{3}{2} A^{2} K_{ij} K^{ij} 
\nonumber \\
&& \quad
  - \bar{\nabla}_i \nu \bar{\nabla}^i \nu,  
\label{e:Einstein4}
\end{eqnarray}
where we introduced the auxiliary functions
\begin{equation}
\nu = \ln N, \quad \zeta =\ln (N A), \quad \beta = \ln B.
\end{equation}
Note that $\bar{\nabla}_i$ denotes the covariant derivative in terms
of a flat 3-metric, $\bar{\triangle} = \bar{\nabla}_i \bar{\nabla}^i$
the corresponding Laplacian, $\Delta_2$ is the 2-dimensional
Laplacian, $U^i$ is the spatial component of 3-velocity
\begin{eqnarray}
&&   \Delta_2  =
    \ddps{}{r} + {1 \over r}\ddp{}{r} +{1 \over r^2}\ddps{}{\theta},
\\
&& U^{r} = - \frac{N^r}{N},
U^{\theta} = - \frac{N^{\theta}}{N},
U^{\varphi} = \frac{1}{N} (\Omega - N^{\varphi})
.
\end{eqnarray}
In Eqs.~(\ref{e:Einstein1}), (\ref{e:Einstein2}), (\ref{e:Einstein3}),
and (\ref{e:Einstein4}), $E$ and $U^i$ are the energy density and the
3-velocity, measured by the locally non-rotating observer: $E =
\gamma^2 [\rho (1 + \varepsilon) + P)]- P$, $\gamma = [1 - [A^{2}
    [(U^r)^2 + r^2 (U^{\varphi})^2] + r^2 \sin^2 \theta B^2
    (U^{\varphi})^2]]^{-1/2}$, where $\rho$ is the comoving rest-mass
density, $\varepsilon$ the specific internal energy, $P$ the
pressure. 

The equation for the shift,
\begin{eqnarray}
\bar{\triangle} N^{i} + 
\frac{1}{3} \bar{\nabla}^{i} \bar{\nabla}_{j} N^{j} = J^i ,
\label{shifteq}
\end{eqnarray}
can be further simplified by introducing a vector $W_{i}$ and a scalar
$\chi$ according to
\begin{eqnarray}
\triangle W_{i} &=& J_{i},\\
\triangle \chi &=& - J_{i} x^{i}. 
\end{eqnarray}
The shift can then be computed from
\begin{equation}
N^{i} = \frac{7}{8} W^{i} - 
\frac{1}{8} [\bar{\nabla}^{i} \chi + \bar{\nabla}^{i} (W_{k} x^{k})],
\end{equation}
and will automatically satisfy Eq.~(\ref{shifteq}).  The vector-type
Poisson equation [Eq.~(\ref{shifteq})] for $N^i$ has hence been
reduced to four scalar-type Poisson equations for $W^i$ and $\chi$.

Let us now discuss the matter part.  We treat the matter as a perfect
fluid, the energy momentum tensor of which is
\begin{equation}
T^{\mu \nu} = \rho \left( 1 + \varepsilon + \frac{P}{\rho} \right)
u^{\mu} u^{\nu} + P g^{\mu \nu},
\end{equation}
where $u^{\mu}$ is the fluid 4-velocity.  We adopt a $\Gamma$-law
equation of state in the form
\begin{equation}
P = (\Gamma - 1) \rho \varepsilon
,
\label{eqn:Gamma-law}
\end{equation}
where $\Gamma$ is the adiabatic index.  In the absence of thermal
dissipation, Eq.~(\ref{eqn:Gamma-law}) together with the first law of
thermodynamics imposes a polytropic equation of state
\begin{equation}
P = \kappa \rho^{1+1/n}
,
\end{equation}
where $n = 1/(\Gamma - 1)$ is the polytropic index and $\kappa$ a
constant.

The relativistic Euler equation is described in axisymmetric
stationary spacetime as 
\begin{equation}
\frac{h_{,j}}{h} - \frac{u^{t}_{,j}}{u^{t}} + u^{t} u_{\varphi}
\Omega_{,j} = 0 \quad (j = r, \theta ),
\label{eqn:REuler}
\end{equation}
where $h \equiv ( 1 + \varepsilon + P/\rho )$ is the specific
enthalpy.  The Bernoulli's equation is derived by integrating the
relativistic Euler equation [Eq.~(\ref{eqn:REuler})] as (see
e.g. \citep{Gourg06})
\begin{equation}
H - K + R_{\rm rot} = C,
\label{eqn:Bernoulli}
\end{equation}
where 
\begin{eqnarray}
H &\equiv& \int \frac{dh}{h} = \ln h = \ln \left[ 1 + \Gamma
  \varepsilon \right]
,\\
K &\equiv& \int \frac{du^{t}}{u^{t}} = \ln u^{t} = -\nu  - \ln \gamma 
,
\end{eqnarray}
and $C$ is a constant of integration.  We adopt two types of rotation
profile, uniform and differential rotation, in this paper.  For a
uniformly rotating star, we simply set the rotational energy potential
$R_{\rm rot}=0$ since $\Omega_{,j}=0$.  For a differentially rotating
star, we assume a specific type of rotation law as 
\begin{equation}
u^{t} u_{\varphi} = A_{\rm rot}^{2} (\Omega_{\rm c} - \Omega)
,
\label{eqn:RLaw}
\end{equation}
where $\Omega_{c}$ is the central angular velocity, $A_{\rm rot}$ the
degree of differential rotation, in order to integrate
Eq.~(\ref{eqn:REuler}) analytically.  In the Newtonian limit ($u^{t}
\rightarrow 1$ and $u^{\varphi} \rightarrow \varpi^{2} \Omega$, where
$\varpi$ is the cylindrical radius), the corresponding rotational
profile reduces to
\begin{equation}
\Omega = \frac{A_{\rm rot}^{2} \Omega_c}{\varpi^{2} + A_{\rm rot}^{2}},
\label{eqn:RLaw_Newton}
\end{equation}
which is called $j$-constant rotation law.  The rotational potential
energy for this case is
\begin{equation}
R_{\rm rot} \equiv \int u^{t} u_{\varphi} d\Omega 
= -\frac{1}{2} A_{\rm rot}^{2} (\Omega_{c} - \Omega)^{2}
.
\end{equation}
The enthalpy is derived from the Bernoulli's equation
[Eq.~(\ref{eqn:Bernoulli})] as
\begin{eqnarray}
H &=& H^{\rm max} + (K - K^{\rm max}) - (R_{\rm rot} - R^{\rm
  max}_{\rm rot})
\nonumber \\
&=& \ln(1 + \Gamma \varepsilon^{\rm max}) + (\nu^{\rm max} - \nu)  +
(\ln \gamma^{\rm max} - \ln \gamma)
\nonumber \\
&&
- (R_{\rm rot} - R^{\rm max}_{\rm rot})
,
\end{eqnarray}
where $H^{\rm max}$, $K^{\rm max}$, $R^{\rm max}_{\rm rot}$ represents
the values at the maximum enthalpy.  The rotation law of the star
[Eq.~(\ref{eqn:RLaw})] becomes
\begin{equation}
A_{\rm rot}^{2} (\Omega_{c} - \Omega) = \frac{r^2 \sin^2 \theta B^2
  (\Omega - N^{\varphi})}{N^2 - r^2 \sin^2 \theta B^2 (\Omega -
  N^{\varphi})^2}
.
\end{equation}

We also rescale the gravitational constant during the iteration
process in order to determine the radius of the star \citep{BGSM93}.
Since the gravitational constant only enters through the matter of the
star, we split the equation for lapse into two parts: One comes from
the spacetime geometry and the other from the matter, and solve them
independently ($\nu = \nu_{\rm q} + \nu_{\rm m}$, $\nu_{\rm q}$ is the
contribution from the spacetime geometry and $\nu_{\rm m}$ the one
from the matter).  After that we set the equatorial surface in the
computational domain and varies the gravitational constant
\citep{footnote} in the following:
\begin{widetext}
\begin{equation}
G = 
  \frac{(H^{\rm max} + \nu_{\rm q}^{\rm max} 
    - \ln\gamma^{\rm max} + R^{\rm max}_{\rm rot}) - 
    (H^{\rm sur} + \nu_{\rm q}^{\rm sur} 
    - \ln\gamma^{\rm sur} + R^{\rm sur}_{\rm rot})
  }
  {\nu_{\rm m}^{\rm sur} - \nu_{\rm m}^{\rm max}}
,
\end{equation}
\end{widetext}
where $H^{\rm max}$, $\nu_{\rm q}^{\rm max}$, $\ln\gamma^{\rm max}$,
$R^{\rm max}$ denotes the value at the maximum enthalpy, $H^{\rm
  sur}$, $\nu_{\rm q}^{\rm sur}$, $\ln\gamma^{\rm sur}$, $R^{\rm sur}$
denotes the value at the equatorial surface of the star, which are
unknown in each iteration step.

In order to test our numerical code internally, we check the virial
identities GRV2 \citep{BG94} and GRV3 \citep{GB94}, the latter being a
relativistic version of the classical virial theorem. The relative
errors are defined by
\begin{widetext}
\begin{eqnarray}
{\rm GRV2} &=& 1 + \frac{
   \int_{0}^{\pi}d\theta  \int_{0}^{\infty} rdr ~
     \sigma_{\rm GRV2m} (r,\theta)} 
  {\int_{0}^{\pi}d\theta \int_{0}^{\infty} rdr ~
     \sigma_{\rm GRV2q} (r,\theta)},
\\
{\rm GRV3} &=& \frac{
  \int_{0}^{2\pi} d\varphi ~
  \int_{0}^{\pi} \sin\theta d\theta ~
  \int_{0}^{\infty} r^2dr ~
    [\sigma_{\rm GRV3q} (r,\theta, \varphi) + 
     \sigma_{\rm GRV3m} (r,\theta, \varphi)]}
  {\int_{0}^{2\pi} d\varphi
   \int_{0}^{\pi} \sin\theta d\theta
   \int_{0}^{\infty} r^2dr ~
    \sigma_{\rm GRV3m} (r,\theta,\varphi) }
.
\end{eqnarray}
\end{widetext}
where
\begin{eqnarray}
\sigma_{\rm GRV2m} (r,\theta) &=&
  8 \pi A^2 [P + (E + P) U_i U^i]
,\\
\sigma_{\rm GRV2q} (r,\theta) &=&
  \frac{3}{2} A^2 K_{ij} K^{ij} - \bar{\nabla}_i \nu \bar{\nabla}^i
  \nu
,\\
\sigma_{\rm GRV3m} (r,\theta,\varphi) &=&
  4 \pi A^2 B [3 P + (E + P) U_i U^i]
, \nonumber \\
&& \\
\sigma_{\rm GRV3q} (r,\theta,\varphi) &=&
  \frac{3}{4} A^2 K_{ij} K^{ij}  - \bar{\nabla}_i \nu \bar{\nabla}^i
  \nu 
\nonumber \\
&&
  + \frac{1}{2} \bar{\nabla}_i \alpha \bar{\nabla}^i \beta
.
\end{eqnarray}
Note that the two quantities GRV2 and GRV3, that should be identically
0 in the ideal equilibrium configuration, have already been rescaled
by the typical source term of the equation, and therefore
automatically defined as a relative error of our computation.

The gravitational mass $M$, proper mass $M_p$, total angular momentum
$J$, rotational kinetic energy $T$, gravitational binding energy $W$
can be computed from
\begin{eqnarray}
M &=& 
  \int_{0}^{2\pi} d\varphi 
  \int_{0}^{\pi} \sin\theta d\theta
  \int_{0}^{\infty} r^2dr~
  A^2 B [N [E + 3 P 
\nonumber \\
&& + (E+P) U_i U^i]
  + 2 r \sin \theta B (E + P) N^{\varphi} U_i U^i],
\nonumber \\
&&\\
M_p &=& 
  \int_{0}^{2\pi} d\varphi 
  \int_{0}^{\pi} \sin\theta d\theta
  \int_{0}^{\infty} r^2dr~
  A^2 B \gamma \rho
,\\
J &=&
  \int_{0}^{2\pi} d\varphi 
  \int_{0}^{\pi} \sin\theta d\theta
  \int_{0}^{\infty} r^2dr
\nonumber \\
&&
  \times r \sin \theta A^2 B^2 (E + P)  U_i U^i
,\\
T &=& \frac{1}{2}
  \int_{0}^{2\pi} d\varphi 
  \int_{0}^{\pi} \sin\theta d\theta
  \int_{0}^{\infty} r^2dr
\nonumber \\
&&
  \times r \sin\theta \Omega A^2 B^2 (E + P) U_i U^i
,\\
W &=& M_p + T - M
.
\end{eqnarray}

Since we use a polytropic equation of state in the equilibrium, it is
convenient to rescale all quantities with respect to $\kappa$.  Since
$\kappa^{n/2}$ has dimensions of length, we introduce the following
nondimensional variables
\begin{equation}
\begin{array}{c c c}
\bar{M} = \kappa^{-n/2} M
, &
\bar{R} = \kappa^{-n/2} R
, &
\bar{J} = \kappa^{-n} J
,
\\
\bar{T} = \kappa^{-n/2} T
, &
\bar{W} = \kappa^{-n/2} W
, &
\bar{\Omega} = \kappa^{n/2} \Omega
. 
\end{array}
\end{equation}
Henceforth, we adopt nondimensional quantities, but omit the bars for
convenience (equivalently, we set $\kappa = 1$).

\begin{figure*}
\centering
\includegraphics[keepaspectratio=true,width=10cm]{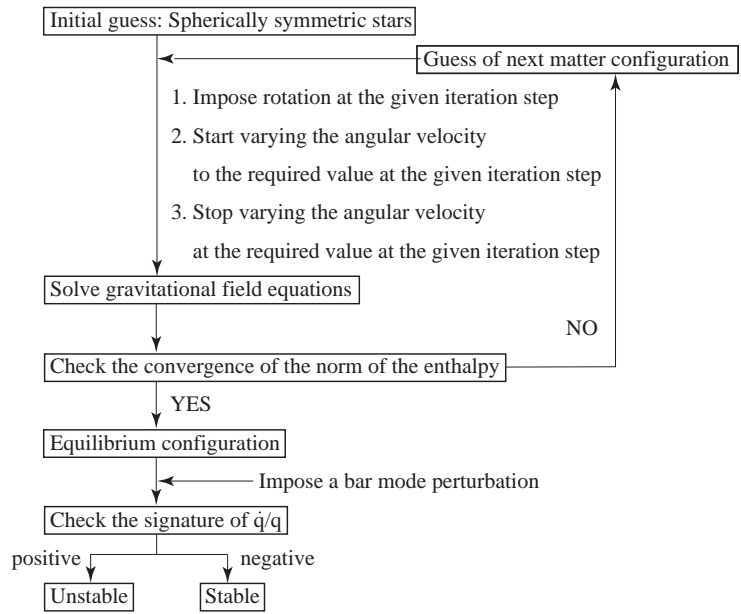}
\caption{
Sketch of the computational procedure to determine the stability of
rotating stars.
}
\label{fig:scheme}
\end{figure*}

Our computations have been made via a multi-domain spectral method
\citep{BGM98}.  We have developed a code to implement this method by
using the C++ library {\sc Lorene} \citep{lorene}.  The key advantage
of the spectral method is that the required number of grid points to
obtain a sufficiently high accuracy is quite small compared to the
grid points in finite differencing, and the accuracy is guaranteed up
to the round-off error in principle.  Since this method is only
applicable to smooth functions, we treat the discontinuity at the
surface of the star by splitting the computational domain.  Note that
we introduce three computational domains to cover the space.  The
innermost domain covers the whole star, while the outermost one is
compactified which allows to cover the space up to spatial infinity.
We also use surface fitting method to split the domain at the surface
of the star \citep{BGM98}.  This method works perfect until the
equatorial surface of the star has a cusp, which happens when the
uniformly rotating star approaches to the mass shedding limit, or the
star is highly deformed from the sphere due to differential rotation. 

\subsection{Iterative evolution approach}
We follow the iterative evolution approach \citep{BFG96, BFG98} to
investigate the viscosity driven instability in rotating relativistic
stars.  We particularly focus on the effect of relativistic gravitation
for compressible fluids.  The physical viewpoint of this approach is
only shown in Newtonian incompressible star that to study the
transition from a uniformly rotating axisymmetric body (Maclaurin
spheroid) to a nonaxisymmetric body (Jacobi ellipsoid). According to
\citet{CKST95}, the above deformation process is driven by viscosity,
since it only varies the circulation but keeps the other two conserved
quantities, total energy and angular momentum, in the Newtonian
incompressible star.  The computational viewpoint of this approach is
that instead of performing the time evolution of the star to
investigate the stability of the star, we treat the iterative number
as evolutional time and determine the stability of the star.  The
advantage of this approach is that there 
is no restriction to the evolutional timestep even in a high
compactness star.  Note that it is entirely difficult to study the
secular instability in the explicit evolution scheme in full general
relativistic hydrodynamics, since the restriction from the Courant
timestep scales as $\sim (M/R)^{-1/2}$.  The uncertain issue in this
approach is that whether one can treat iterative number as evolutional
time, and there is no relationship between each other in a
mathematical sense.  However there is a correspondence: In the
Newtonian incompressible star, \citet{GG02} investigate the difference
between the exact critical value on the bifurcation point
(e.g. \citep{Chandra69}) and find that it is within the round-off
error.

Our computational study of the iterative evolution approach is divided
into two stages; construction of a rotating equilibrium star and the
determination of the viscosity driven bar mode stability of a rotating
equilibrium star.  To construct a rotating equilibrium star, we first
construct a spherical star with given parameters of $\Gamma$ and
$H_{\rm max}$, as an initial guess of the metric components and the
matter profile.  Next we solve gravitational field equations and
determine the new matter profile for the next iteration step.  During
the iteration we impose rotation at the given iteration step, start
varying the angular velocity to the given required value at the given
iteration step, stop varying the angular velocity at the given
required value at the given iteration step.  We stop our iteration
cycle when the relative error of the enthalpy norm between the
previous step and the current step is within $1.0 \times 10^{-7}$.  We
check the relativistic virial identity and the identity for all the
equilibrium configuration and find that the relative errors of GRV2
and GRV3 are $\lesssim$ several $\times 10^{-4}$.

To determine the stability of rotating relativistic star driven by
viscosity, we follow all the computational procedure to construct a
rotating equilibrium star until we reach the relative error of the
enthalpy norm at $1.5 \times 10^{-7}$.  At this iteration step, we put
the following $m=2$ perturbation in the logarithmic lapse to enhance
the growth of the bar mode instability as
\begin{equation}
\nu = \nu_{\rm eq} 
  ( 1 + \varepsilon_{\rm amp}\sin^{2}\theta \cos 2 \varphi ),
\end{equation}
where $\nu_{\rm eq}$ is the logarithmic lapse in the equilibrium,
$\varepsilon_{\rm amp}$ is the amplitude of the perturbation.  We
diagnose the maximum logarithmic lapse of the $m=2$ coefficients in
terms of mode decomposition as
\begin{equation}
q = {\rm max} ~ | \hat{\nu}_{2} |,
\end{equation}
where
\begin{equation}
\nu = \sum_{m=0}^{\infty} \hat{\nu}_{m} e^{i m \varphi}.
\end{equation}
We also define the logarithmic derivative of $q$ in the iteration step
${\cal N}_i$ as
\begin{equation}
\frac{\dot{q}}{q} = \frac{q_{i} - q_{i-1}}{q_{i-1}},
\end{equation}
where $q_i$ denotes $q$ at the iteration step ${\cal N}_i$.  We
determine the stability of the star in terms of viscosity driven
instability as follows.  When the diagnostic $q$ grows exponentially
after we impose a bar mode perturbation in the logarithmic lapse, we
conclude that the star is unstable.  On the other hand when the diagnostic
decays after we put a perturbation, the star is stable.
More precisely, we monitor the derivative of $q$ after we impose a
bar-mode perturbation.  We conclude that the equilibrium star is unstable
when the $\dot{q}/q$ settles down to a positive constant value, while
stable to a negative constant value.  Note that the existence of the
plateau in $\dot{q}/q$ after the several iteration steps once we put a 
perturbation confirms us that we are in the linear perturbation
regime, and therefore guarantees our choice of the perturbation
amplitude we imposed ($\varepsilon_{\rm amp}$).  Finally we determine
the critical value of $T/W$ as the minimum one in the unstable
branch.  We also confirm our argument in all equilibrium stars that
there is a continuous transition between stable and unstable stars as
a function of $T/W$ in our model.  We summarize our computational
procedure in Fig.~\ref{fig:scheme}.

\section{Uniformly rotating stars}
\label{sec:rigid}

\begin{figure}
\centering
\includegraphics[keepaspectratio=true,width=9cm]{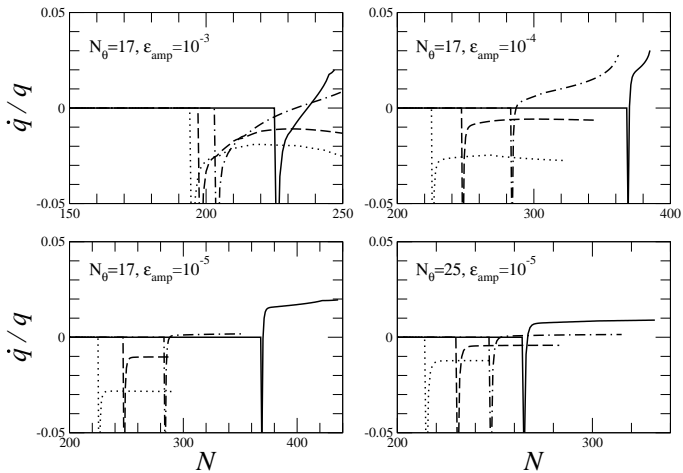}
\caption{
Diagnostic $\dot{q}/q$ as a function of iteration numbers ${\cal N}$ 
for sixteen different uniformly rotating stars of $\Gamma=2.3$ and
$M/R=0.01$.  Solid, dash-dotted, dashed, dotted lines denotes  
$T/W = (0.1306, 0.1307, 0.1308, 0.1309)$ for 
$({\cal N}_{\theta}, \varepsilon_{\rm amp}) = (17, 1.00 \times 10^{-3})$, 
$T/W = (0.1309, 0.1312, 0.1313, 0.1314)$ for 
$({\cal N}_{\theta}, \varepsilon_{\rm amp}) = (17, 1.00 \times 10^{-4})$, 
$T/W = (0.1309, 0.1312, 0.1313 , 0.1314)$ for 
$({\cal N}_{\theta}, \varepsilon_{\rm amp}) = (17, 1.00 \times 10^{-5})$, 
and 
$T/W = (0.1306, 0.1307, 0.1308 , 0.1310)$ for 
$({\cal N}_{\theta}, \varepsilon_{\rm amp}) = (25, 1.00 \times 10^{-5})$, 
respectively.
}
\label{fig:testdig}
\end{figure}

\begin{figure}
\centering
\includegraphics[keepaspectratio=true,width=8cm]{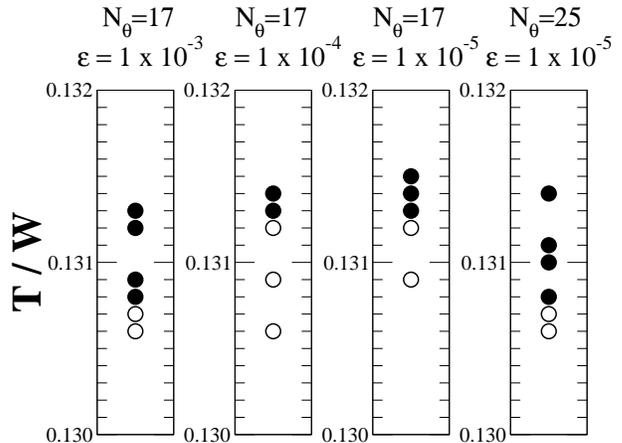}
\caption{
Stability of uniformly rotating stars of  $\Gamma = 2.3$, $M/R=0.01$
for four different parameters.  Circle (open) and circle (filled)
denotes stable and unstable to viscosity driven bar mode perturbation
respectively.  We fix the compactness for each parameter up to four
digits.  Note that there is a monotonic transition from stable to
unstable when increasing $T/W$.
}
\label{fig:testtw}
\end{figure}

\begin{figure*}
\centering
\includegraphics[keepaspectratio=true,width=16cm]{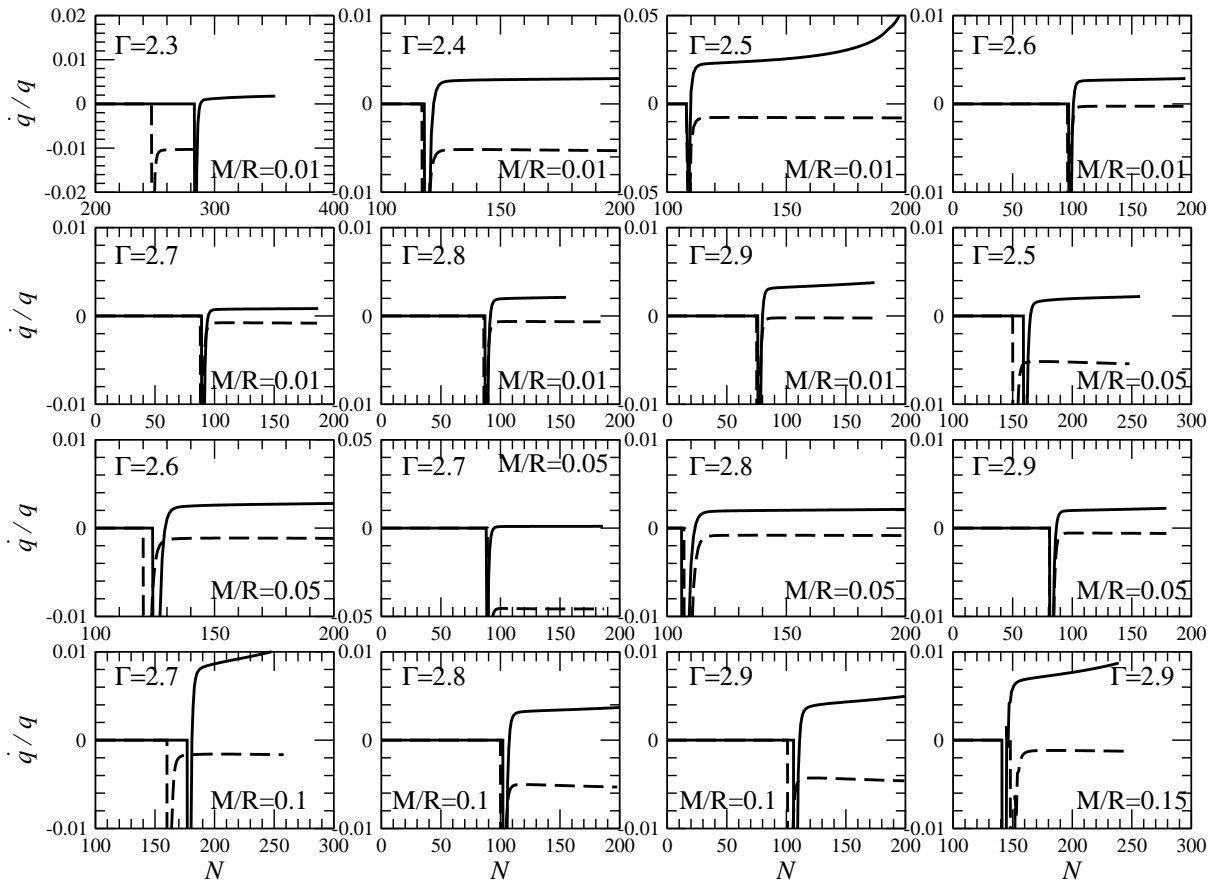}
\caption{
Diagnostic $\dot{q}/q$ as a function of iteration steps ${\cal N}$ for
sixteen different uniformly rotating stars.  Solid and dashed lines
denotes the unstable and stable stars respectively.  Note that the
$T/W$ for each stable star in the same compactness is $0.0001$ lower
than the critical value of that of an unstable star (see
Table~\ref{tab:rigid}).
}
\label{fig:rigiddig}
\end{figure*}

Before studying the viscosity driven instability in uniformly rotating
stars, we examine the dependence of the perturbation amplitude
$\varepsilon_{\rm amp}$ and the collocation points on the critical
value of viscosity driven bar mode instability $(T/W)_{\rm crt}$.
Note that we introduce three domains to cover the whole
space, and each domain has a relationship of ${\cal N}_r = 2 {\cal
  N}_{\theta} - 1$ and ${\cal N}_{\varphi} = 4$, where ${\cal N}_r$,
${\cal N}_{\theta}$, ${\cal N}_{\varphi}$ represents the collocation
points for the radial direction, the polar angle, the azimuthal angle,
respectively.

First we vary the amplitude $\varepsilon_{\rm amp}$ in the range
between $1.00 \times 10^{-3}$ -- $1.00 \times 10^{-5}$.  We show the
diagnostic $\dot{q}/q$ in Fig.~\ref{fig:testdig}.  The general picture
of $\dot{q}/q$ in our computation is composed of three stages: (1)
$\dot{q}/q=0$ (2) Sudden change in $\dot{q}/q$ at a certain iteration
step (3) Continuous smooth function.  (1) represents the stage of
constructing the axisymmetric rotating equilibrium
configuration.  Since we solve the equations in the axisymmetric
spacetime, there is no non-axisymmetric contribution in this stage and
therefore both $q$ and $\dot{q}/q$ are $0$ up to numerical error.  (2)
represents a reaction due to a sudden imposition of a bar-mode
perturbation in the logarithmic lapse.  The quantity $\dot{q}/q$
should drastically change at this iteration step.  (3) represents the
post-perturbation stage, which determines the stability of the
equilibrium star.  When $\dot{q}/q$ takes a positive value after a
certain iteration step we conclude that the equilibrium star is unstable,
while takes a negative value stable.  Note that it is in the linear
perturbation regime when $\dot{q}/q$ takes a constant value after the
perturbation.  Therefore the solid and dash-dotted lines in
Fig.~\ref{fig:testdig} are stable while dash and dotted lines are
unstable (plotted in Fig.~\ref{fig:testtw}).  The diagnostic
$\dot{q}/q$ suggests us to impose the amplitude below $1.00 \times
10^{-4}$, since there is no plateau for the case of ${\cal N}_{\theta}
= 17$, $\varepsilon_{\rm amp} = 1.00 \times 10^{-3}$ in the stable star
(Fig.~\ref{fig:testdig}).  We also check whether we have a continuous
transition from the stable star to the unstable one in terms of $T/W$
(Fig.~\ref{fig:testtw}), and confirmed that there exists a minimum
$T/W$ in the unstable stars.  We determine the minimum value as
$(T/W)_{\rm crt}$.  The monotonic increase of the final value of
$\dot{q}/q$ as increasing $T/W$ (Fig.~\ref{fig:testdig}) also supports
the previous statement.  We find in Table~\ref{tab:test} that the
critical value of $T/W$ depends on the choice of $\varepsilon_{\rm
  amp}$ for only $0.4\%$, which means that we are in the high
convergence level of the choice of $\varepsilon_{\rm amp}$.

Next we show our result of two different choices of the collocation
points ${\cal N}_{\theta} = 17$ and $25$ in Table~\ref{tab:test}, and
find that the critical $T/W$ depends on the choice of ${\cal N}_r$ and
${\cal N}_{\theta}$ only for $0.4\%$.  This means that we are also in
the high convergence level in the choice of collocation
points. Therefore we briefly estimate that our accuracy level of the
critical value of $T/W$ is $\lesssim 1\%$.  Hereafter we choose the
parameter sets $\varepsilon_{\rm amp} = 1.00 \times 10^{-5}$ and
${\cal N}_{\theta} = 17$ to determine the critical value of viscosity
driven instability in uniformly rotating stars.

\begin{table}[htbp]
\begin{center}
\leavevmode
\caption{
Dependence of critical $T/W$ on the amplitude $\varepsilon_{\rm amp}$
and on the collocation points in uniformly rotating stars of $\Gamma =
2.3$, $M/R=0.01$.
}
\begin{tabular}{c c c}
\hline
\hline
$\varepsilon_{\rm amp}$ & ${\cal N}_{\theta}$ & $(T/W)_{\rm crt}$
\\
\hline
$1.00 \times 10^{-3}$ & $17$ & $0.1308$
\\
$1.00 \times 10^{-4}$ & $17$ & $0.1313$
\\
$1.00 \times 10^{-5}$ & $17$ & $0.1313$
\\
\hline
$1.00 \times 10^{-5}$ & $25$ & $0.1308$
\\
\hline
\hline
\end{tabular}
\label{tab:test}
\end{center}
\end{table}

\begin{figure*}
\centering
\includegraphics[keepaspectratio=true,width=16cm]{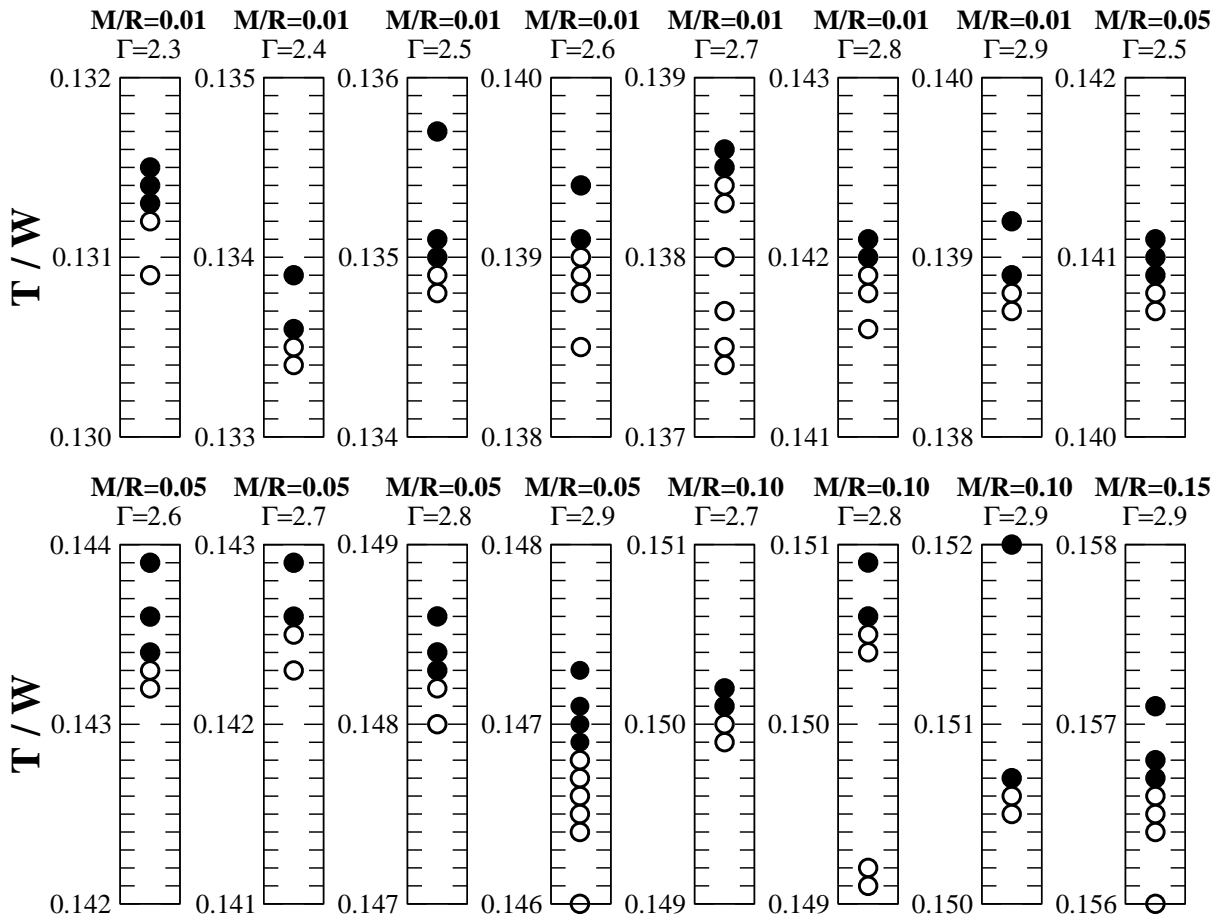}
\caption{
Stability of uniformly rotating stars for sixteen different
parameters.  Circle (open) and circle (filled) denotes stable and
unstable to viscosity driven bar mode perturbation respectively.  We
fix the compactness for each parameter up to four digits.  Note that
there is a monotonic transition from stable to unstable when
increasing $T/W$.
}
\label{fig:rigidtw}
\end{figure*}

\begin{figure}
\centering
\includegraphics[keepaspectratio=true,width=8cm]{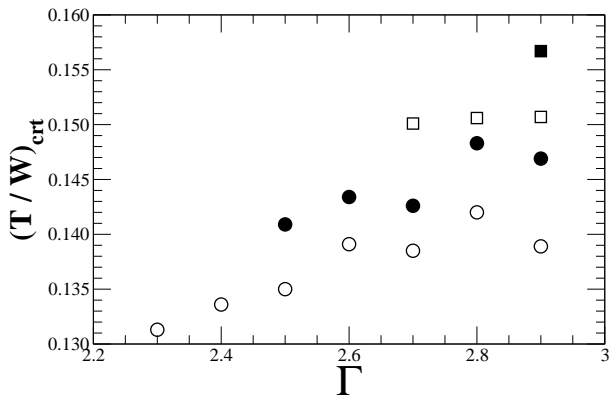}
\caption{
Critical value of $T/W$ as a function of $\Gamma$ for four different
compactness of uniformly rotating stars (see
Table~\ref{tab:rigid}).  Circle (open), circle (filled), square
(open), and square (filled) denotes the compactness ($M/R$) of $0.01$,
$0.05$, $0.1$, and $0.15$, respectively.  The star whose adiabatic
index is $\Gamma_{\rm low} - 0.1$, where $\Gamma_{\rm low}$ is the
lowest $\Gamma$ of the unstable star in each compactness in the
figure, is stable.
}
\label{fig:rigid}
\end{figure}

After we determine our choice of $\varepsilon_{\rm amp}$ and the
collocation points, we study the critical value of $T/W$ of the
viscosity driven instability in uniformly rotating stars.  We show the
diagnostic of the two closest star to the critical value of
$T/W$ for sixteen different parameters in Fig.~\ref{fig:rigiddig}.
Note that there 
is a clear plateau after we put a perturbation, and therefore we are
in the linear perturbation regime.  We also study the stability of
the stars in terms of $T/W$ (Fig.~\ref{fig:rigidtw}) and determine the
critical values of instability, which are summarized in
Table~\ref{tab:rigid} and in Fig.~\ref{fig:rigid}.  We find that the
relativistic gravitation stabilizes the star from viscosity driven
instability, and that the critical value of $T/W$ for each compactness
is almost insensitive to the polytropic $\Gamma$ of the equation of
state.  The Newtonian compressible calculation has been performed in
Fig.~3 of \citet{BFG96} that the critical value of $T/W$ is $\sim
0.134$ which is not so sensitive to the variation of the polytropic
$\Gamma$.  Our computational results (Fig.~\ref{fig:rigid}) shows that
the critical $T/W$ is $\sim 0.137$ for $M/R=0.01$ , $\sim 0.145$ for
$M/R=0.05$, $\sim 0.150$ for $M/R=0.1$, $\sim 0.157$ for $M/R=0.15$,
respectively.  The critical value of $T/W$ monotonically increases
when increasing the compactness of the star, which means that the
relativistic gravitation stabilizes the viscosity driven instability.
Note that the $\Gamma$ below the smallest $\Gamma$ in each compactness
of the star plotted in Fig.~\ref{fig:rigid} represents that the star
is stable up to the mass-shedding limit.

\begin{table*}[htbp]
\begin{center}
\leavevmode
\caption{
Critical value of $T/W$ of viscosity driven instability in uniformly
rotating relativistic stars} 
\begin{tabular}{c c c c c c c c c c}
\hline
\hline
$\Gamma$\footnotemark[1] &
$R_{p}/R_{e}$\footnotemark[2] &
$H_{\rm max}$\footnotemark[3] & 
$R$\footnotemark[4] &
$M$\footnotemark[5] &
$J$\footnotemark[6] &
$(T / W)_{\rm crt}$ &
$M / R$ &
GRV2 & GRV3
\\
\hline
$2.30$ & $0.5276$ & $9.199 \times 10^{-3}$ & $0.9323$ & $0.009322$ &
$0.0002082$ & $0.1313$ & $0.009999$ & $6.44 \times 10^{-5}$ & 
$-1.54 \times 10^{-4}$ 
\\
$2.40$ & $0.5446$ & $8.465 \times 10^{-3}$ & $0.7518$ & $0.007518$ &
$0.0001400$ & $0.1336$ & $0.01000$  & $3.22 \times 10^{-5}$ & 
$-9.60 \times 10^{-5}$ 
\\
$2.50$ & $0.5526$ & $8.002 \times 10^{-3}$ & $0.6322$ & $0.006322$ &
$0.0001008$ & $0.1350$ & $0.01000$  & $-3.78 \times 10^{-5}$ & 
$1.04 \times 10^{-4}$ 
\\
$2.60$ & $0.5482$ & $7.662 \times 10^{-3}$ & $0.5528$ & $0.005528$ &
$7.904 \times 10^{-5}$ & $0.1391$ & $0.01000$  & $5.04 \times 10^{-5}$
& $-2.14 \times 10^{-5}$
\\
$2.70$ & $0.5590$ & $7.368 \times 10^{-3}$ & $0.4825$ & $0.004825$ &
$6.057 \times 10^{-5}$ & $0.1385$ & $0.01000$  & $7.99 \times 10^{-5}$
& $-1.73 \times 10^{-4}$
\\
$2.80$ & $0.5510$ & $7.130 \times 10^{-3}$ & $0.4352$ & $0.04352$ &
$5.023 \times 10^{-5}$ & $0.1420$ & $0.01000$ & 
$-1.24 \times 10^{-5}$ & $8.59 \times 10^{-5}$
\\
$2.90$ & $0.5663$ & $6.933 \times 10^{-3}$ & $0.3879$ & $0.003879$ &
$3.966 \times 10^{-5}$ & $0.1389$ & $0.01000$  & $5.85 \times 10^{-5}$
& $-1.52 \times 10^{-4}$
\\
\hline
$2.50$ & $0.5312$ & $4.542 \times 10^{-2}$ & $0.8140$ & $0.04071$ &
$0.001986$ & $0.1409$ & $0.05001$ & $4.72 \times 10^{-5}$ & 
$-1.30 \times 10^{-4}$ 
\\
$2.60$ & $0.5368$ & $4.291 \times 10^{-2}$ & $0.7267$ & $0.03633$ &
$0.0001616$ & $0.1434$ & $0.05000$ & $5.04 \times 10^{-5}$ & 
$-4.94 \times 10^{-5}$  
\\
$2.70$ & $0.5486$ & $4.090 \times 10^{-2}$ & $0.6513$ & $0.03256$ &
$0.001306$ & $0.1426$ & $0.05000$ & $7.01 \times 10^{-5}$ & 
$-1.64 \times 10^{-4}$ 
\\
$2.80$ & $0.5380$ & $3.951 \times 10^{-2}$ & $0.6089$ & $0.03044$ &
$0.001171$ & $0.1483$ & $0.05000$ & $-1.02 \times 10^{-4}$ & 
$1.34 \times 10^{-4}$ 
\\
$2.90$ & $0.5471$ & $3.820 \times 10^{-2}$ & $0.5586$ & $0.02793$ &
$0.0009868$ & $0.1469$ & $0.05000$ & $4.99 \times 10^{-5}$ & 
$-1.23 \times 10^{-4}$ 
\\
\hline
$2.70$ & $0.5241$ & $9.710 \times 10^{-2}$ & $0.7320$ & $0.07319$ &
$0.005049$ & $0.1501$ & $0.09999$ & $7.93 \times 10^{-5}$ & 
$-7.16 \times 10^{-5}$ 
\\
$2.80$ & $0.5351$ & $9.197 \times 10^{-2}$ & $0.6734$ & $0.06735$ &
$0.004319$ & $0.1506$ & $0.1000$ & $7.23 \times 10^{-5}$ & 
$-1.86 \times 10^{-4}$ 
\\
$2.90$ & $0.5423$ & $8.824 \times 10^{-2}$ & $0.6274$ & $0.06274$ &
$0.003773$ & $0.1507$ & $0.1000$ & $8.09 \times 10^{-5}$ & 
$-3.63 \times 10^{-5}$ 
\\
\hline
$2.90$ & $0.5242$ & $1.608 \times 10^{-1}$ & $0.6634$ & $0.09957$ &
$0.008380$ & $0.1567$ & $0.1501$ & $3.16 \times 10^{-5}$ & 
$-6.99 \times 10^{-5}$ 
\\
\hline
\hline
\end{tabular}
\label{tab:rigid}
\footnotetext[1]{$\Gamma$: Adiabatic index of the equation of state}
\footnotetext[2]{$R_{p}/R_{e}$: Ratio of the polar proper radius to the
equatorial proper radius}
\footnotetext[3]{$H_{\rm max}$: Maximum enthalpy}
\footnotetext[4]{$R$: Circumferential radius}
\footnotetext[5]{$M$: Gravitational mass}
\footnotetext[6]{$J$: Total angular momentum}
\end{center}
\end{table*}

\section{Differentially rotating stars}
\label{sec:diff}
Next we follow the same approach to study the critical value of
viscosity driven bar-mode instability in differentially rotating
stars.  We study two cases based on the variation of rotation profile
during the iteration for differentially rotating stars.  One is that
we fix the rotation profile throughout the iteration.  The physical
timestep which corresponds to the iteration step is much shorter than
the dynamical time in this case since the different fragments of the
fluid in terms of a cylindrical radius move at different azimuthal
speeds, and therefore the trace shows a spiral structure.  The other
is to change the rotation profile throughout the evolution.  Since we
mimic the model that the viscosity changes the rotation profile, the
total angular momentum is approximately conserved throughout the
process.  We use the same collocation points as in the case of
uniformly rotating stars.

\begin{figure*}
\centering
\includegraphics[keepaspectratio=true,width=12cm]{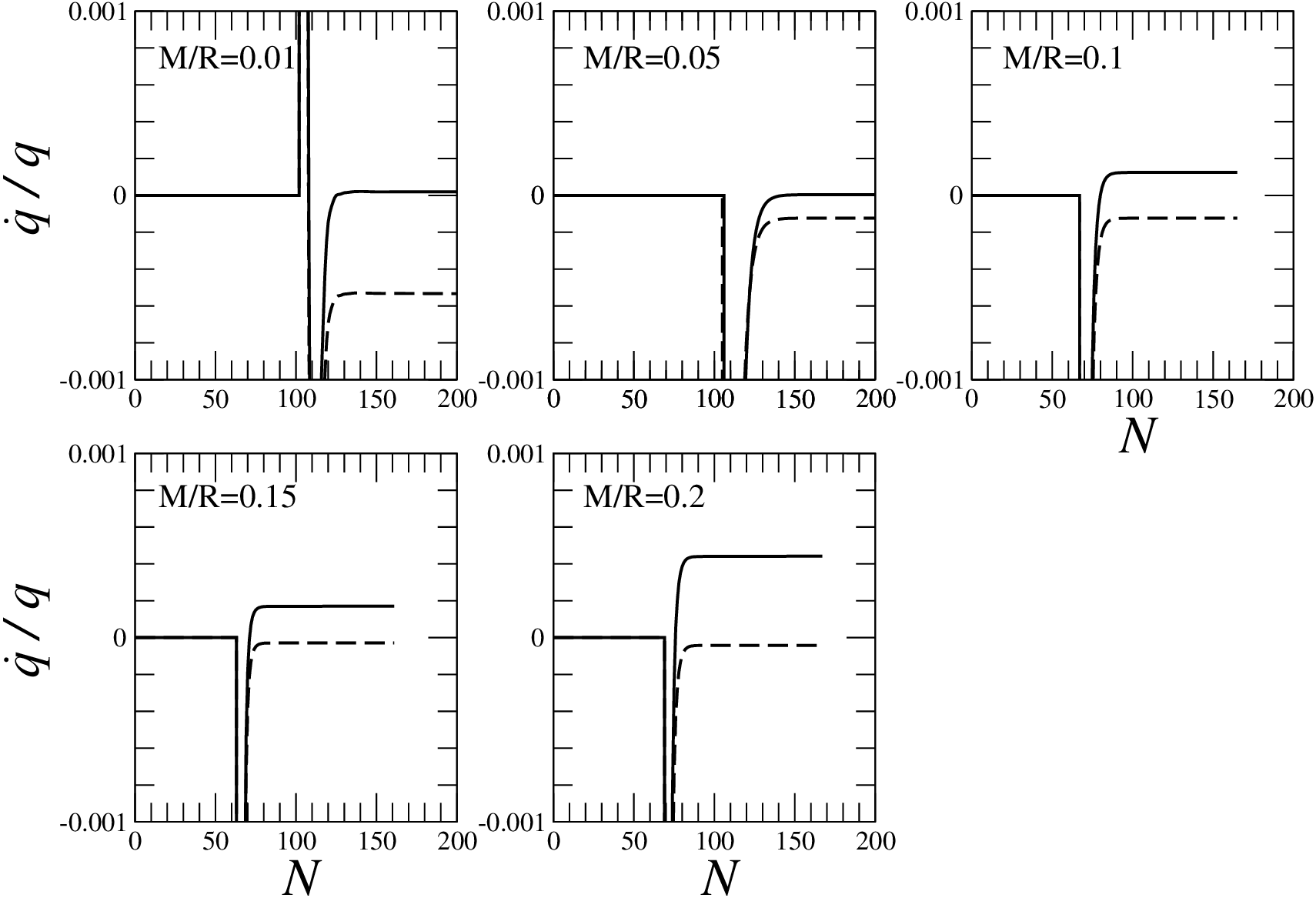}
\caption{
Diagnostic $\dot{q}/q$ as a function of iteration steps ${\cal N}$ for
five different differentially rotating stars.  Solid and dashed lines
denotes the unstable and stable stars respectively.  Note that the
$T/W$ for each stable star in the same compactness is $0.0001$ lower
than the critical value of that of an unstable star (see
Table~\ref{tab:diff100}).
}
\label{fig:diff100}
\end{figure*}

\begin{figure}
\centering
\includegraphics[keepaspectratio=true,width=8cm]{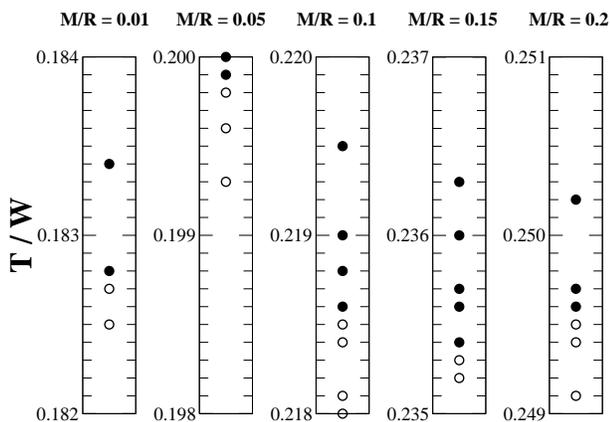}
\caption{
Stability analysis of differentially rotating stars for five different
parameters.  Circle (open) and circle (filled) denotes stable and
unstable to viscosity driven bar mode perturbation respectively.  We
fix the compactness for each parameter up to four digits.  Note that
there is a monotonic transition from stable to unstable when
increasing $T/W$.
}
\label{fig:diff100tw}
\end{figure}

First we show the case of fixed rotation profile throughout the
evolution.  The diagnostic in Fig.~\ref{fig:diff100} shows that there
exists a critical value of viscosity driven instability.  Note that
the plateau at the late stage of the iteration clearly shows that we
are in the linear perturbation regime.  We also monitor the stability
of differentially rotating stars in terms of different $T/W$ in
Fig.~\ref{fig:diff100tw}.  We find a monotonic transition from a
stable star to an unstable star as a function of $T/W$, which
guarantees the determination of critical value of $T/W$ as a minimum
$T/W$ in the unstable stars.  We summarize our finding in
Table~\ref{tab:diff100}.

We find the following two issues in the critical value of $T/W$.  One
is that relativistic gravitation also stabilizes differentially
rotating stars from the viscosity driven instability.  The above
statement also holds in uniformly rotating incompressible relativistic
stars and in uniformly rotating compressible relativistic stars, and
therefore we find that this statement is quite general one.  The other
is that differential rotation also stabilize the star from the
viscosity driven instability.  The critical value of $T/W$ is $0.13
\sim 0.16$ for a uniformly rotating star,
while $0.18 \sim 0.25$ for a differentially rotating
stars with moderate degree of differential rotation.

\begin{figure*}
\centering
\includegraphics[keepaspectratio=true,width=12cm]{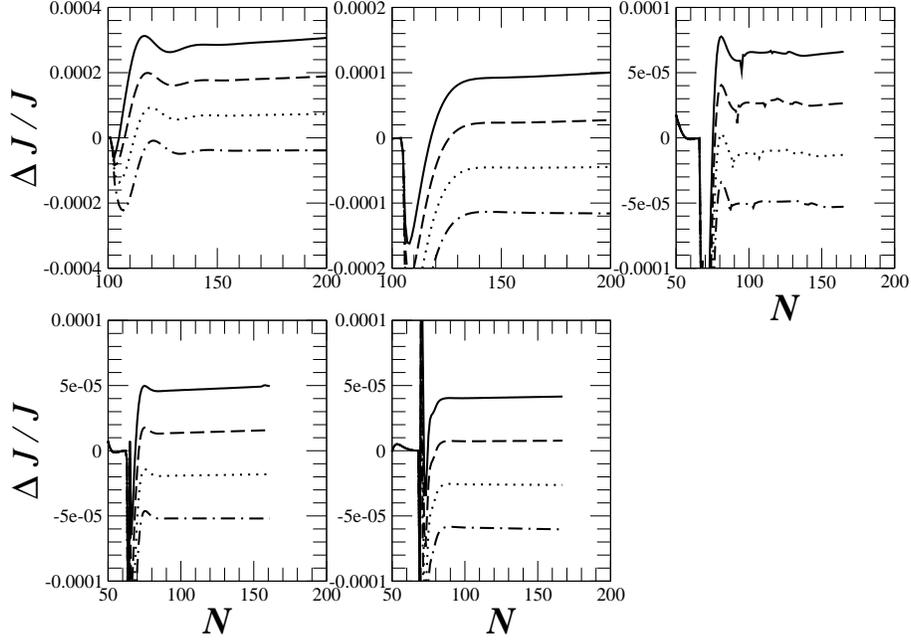}
\caption{
Relative deviation of the total angular momentum between two
successive steps in the iteration process, as a function of the step
number ${\cal N}$.  Solid, dashed, dotted, dash-dotted lines represent 
$\varepsilon_{\rm omg} = (0.5, 0.6, 0.7, 0.8) \times 10^{-4}$ for $M/R
= 0.01$ and $T/W = 0.1828$, 
$\varepsilon_{\rm omg} = (0.8, 0.9, 1.0, 1.1) \times 10^{-4}$ for $M/R
= 0.05$ and $T/W = 0.1999$, 
$\varepsilon_{\rm omg} = (1.1, 1.2, 1.3, 1.4) \times 10^{-4}$ for $M/R
= 0.1$ and $T/W = 0.2186$, 
$\varepsilon_{\rm omg} = (1.0, 1.1, 1.2, 1.3) \times 10^{-4}$ for $M/R
= 0.15$ and $T/W = 0.2354$, and 
$\varepsilon_{\rm omg} = (0.9, 1.0, 1.1, 1.2) \times 10^{-4}$ for $M/R
= 0.2$ and $T/W = 0.2496$, respectively.
}
\label{fig:rela_j}
\end{figure*}

\begin{table*}[htbp]
\begin{center}
\leavevmode
\caption{
Critical value of viscosity driven instability in differentially
rotating relativistic stars.  We choose $\Gamma = 2$ polytropic
equation of state and the degree of differential rotation as
$\hat{A}=1$.
}
\begin{tabular}{c c c c c c c c c c c c}
\hline
\hline
$R_{p}/R_{e}$ &
$H_{\rm max}$ & 
$R$ & $M$ & $J$ &
$(T / W)_{\rm crt}$ & 
$M / R$ &
GRV2 & GRV3 
\\
\hline
$0.4458$ & $7.594 \times 10^{-3}$ & $1.978$ & $0.01978$ & $0.01165$ & 
$0.1828$ & $0.01000$ & $-2.03 \times 10^{-6}$ & $2.28 \times 10^{-6}$ 
\\
$0.3985$ & $3.830 \times 10^{-2}$ & $1.894$ & $0.09468$ & $0.01315$ &
$0.1999$ & $0.05000$ & $3.92 \times 10^{-6}$ & $4.92 \times 10^{-4}$ 
\\
$0.3820$ & $7.492 \times 10^{-2}$ & $1.758$ & $0.1758$ & $0.03565$ &
$0.2186$ & $0.1000$ & $3.88 \times 10^{-7}$ & $-4.52 \times 10^{-6}$ 
\\
$0.3457$ & $1.116 \times 10^{-1}$ & $1.609$ & $0.2414$ & $0.06025$ & 
$0.2354$ & $0.1500$ & $-1.93 \times 10^{-5}$ & $-9.03 \times 10^{-6}$ 
\\
$0.2982$ & $1.581 \times 10^{-1}$ & $1.455$ & $0.2910$ & $0.08210$ &
$0.2496$ & $0.2000$ & $-3.63 \times 10^{-6}$ & $-1.55 \times 10^{-5}$
\\
\hline
\hline
\end{tabular}
\label{tab:diff100}
\end{center}
\end{table*}

Next we study the variation of the rotation profile since viscosity
also takes a significant role to change it.  The timescale between the
change of angular velocity profile and the growth of the viscosity
driven bar-mode are in the same order in low viscosity approximation,
we should check whether our result is significantly affected by the
change of the rotation profile of the star.  In order to mimic this
idea, we vary the parameter of the degree of differential rotation
slightly after we impose a perturbation in the following manner:
\begin{equation}
A_{\rm rot}^{-1} = A_{\rm rot}^{-1~\rm (eq)} 
  [1 - \epsilon_{\rm rot} ({\cal N} - {\cal N}_{\rm ptb})],
\end{equation}
where $\epsilon_{\rm rot}$ is the degree of the variation of the
rotation profile we set to $1.0 \times 10^{-4}$, ${\cal N}$ the
iteration number, ${\cal N}_{\rm ptb}$ the iteration number we impose
the perturbation.  Note that viscosity changes the rotation profile to
the uniform one, we put negative sign in front of $\varepsilon_{\rm
  rot}$.  Since the viscosity only affects the local interaction
between the each fluid components, the total angular momentum is
conserved even the viscosity takes the role.  Therefore we also vary
the central angular velocity $\Omega_{\rm c}$ in a following manner
\begin{equation}
\Omega_{\rm c} = \Omega_{\rm c}^{\rm (eq)} 
  [1 - \epsilon_{\rm omg} ({\cal N} - {\cal N}_{\rm ptb})],
\end{equation}
where $\epsilon_{\rm omg}$ is the degree of the variation of the
central angular velocity in order to maintain the total angular
momentum approximately constant.  Note that we put negative sign in
front of $\epsilon_{\rm omg}$ to play an appropriate role of the
viscosity.  In practice we vary  $\epsilon_{\rm omg}$ in $0.1$ steps
and choose the one that changes the angular momentum minimum at the
given $\varepsilon_{\rm rot}$.  For each case the relative change of
the total angular momentum after we impose the perturbation is in the
order of $\lesssim 10^{-5}$.  We show the relationship between the
relative change of the total angular momentum and $\epsilon_{\rm omg}$
for several cases in Fig.~\ref{fig:rela_j}.

\begin{figure*}
\centering
\includegraphics[keepaspectratio=true,width=12cm]{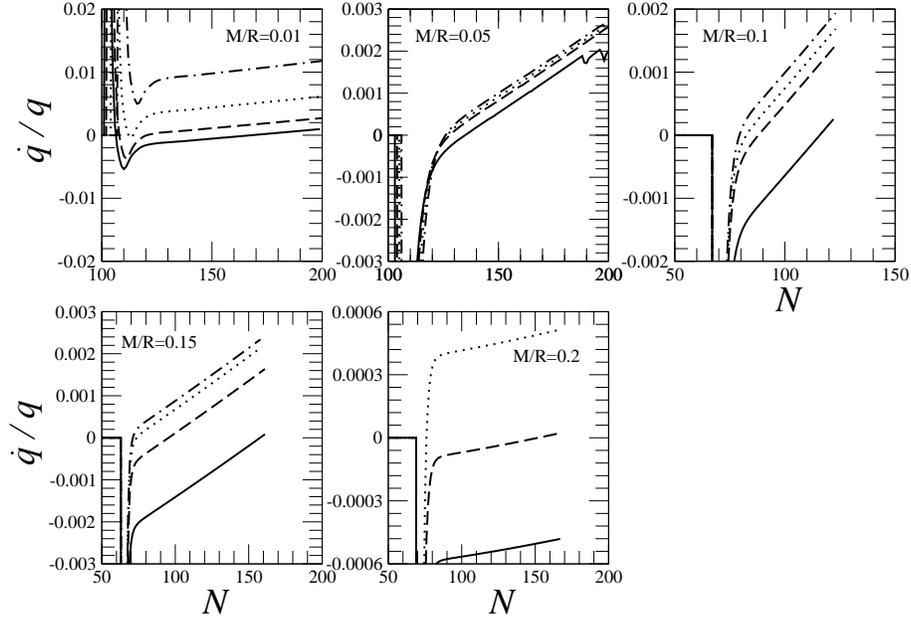}
\caption{
Diagnostic $\dot{q}/q$ as a function of iteration steps ${\cal N}$ for
five different differentially rotating stars.  Solid, dashed, dotted,
and dashed line denotes
$T/W=$ $(0.1825, 0.1828, 0.1834, 0.1844)$ and $\varepsilon_{\rm
  omg}=0.8 \times 10^{-4}$ for $M/R=0.01$,  
$T/W=$ $(0.1993, 0.1996, 0.1998, 0.1999)$ and $\varepsilon_{\rm
  omg}=0.9 \times 10^{-4}$ for $M/R=0.05$,  
$T/W=$ $(0.2180, 0.2184, 0.2185, 0.2186)$ and $\varepsilon_{\rm
  omg}=1.3 \times 10^{-4}$ for $M/R=0.1$,  
$T/W=$ $(0.2348, 0.2352, 0.2353, 0.2354)$ and $\varepsilon_{\rm
  omg}=1.1 \times 10^{-4}$ for $M/R=0.15$, and 
$T/W=$ $(0.2494, 0.2495, 0.2496)$ and $\varepsilon_{\rm omg}=1.0
\times 10^{-4}$ for $M/R=0.2$, 
respectively.  Note that $\dot{q}/q$ is always increasing around the
critical value of $T/W$ in differentially rotating stars, which means
the change of rotational profile due to viscosity unstabilizes the
star.
}
\label{fig:diff100c}
\end{figure*}

Taking into account of the change of rotational profile, we show our
numerical results for the critical values in Fig.~\ref{fig:diff100c}.
We find that all the stars around the critical $T/W$ determined for a
fixed rotation profile becomes unstable.  In fact, the stage of
$\dot{q}/q$ that corresponds to the plateau in Fig.~\ref{fig:diff100}
increases in  Fig.~\ref{fig:diff100c}.  Therefore we estimate the
relevant two timescales, growth of the bar mode due to viscosity and
the variation time of the rotational profile due to viscosity, and
compare them to discuss the condition to induce viscosity driven
instability.

If we assume that the bar grows exponentially throughout the
iteration, the growth timescale of the bar is $(\dot{q}/q)^{-1}$.  The
value of $\dot{q}/q$ at the plateau in Fig.~\ref{fig:diff100}
represent the growth timescale.  On the other hand, the change of the
rotation profile due to viscosity changes the growth timescale of the
bar, and therefore the derivative of the $\dot{q}/q$ represents the
timescale of the change of rotation profile.  We define those two
timescales as
\begin{eqnarray}
\tau_{\rm ang} = (\ddot{q} / q)^{-1/2},
\quad
\tau_{\rm bar} = (\dot{q} / q)^{-1}.
\end{eqnarray}
We summarize those two timescales from our computation in
Table~\ref{tab:timescale}.

\begin{table}[htbp]
\begin{center}
\leavevmode
\caption{
Two timescales in differentially rotating stars extracted from our
numerical results
}
\begin{tabular}{c c c c}
\hline
\hline
$M/R$ & $T/W$ & $\tau_{\rm ang}$ [$\Delta {\cal N}$] & $\tau_{\rm bar}$
    [$\Delta {\cal N}$] 
\\
\hline
$0.01$ & $0.1825$ & $1.8 \times 10^{2}$ & $-6.1 \times 10^{2}$
\\
$0.01$ & $0.1828$ & $1.7 \times 10^{2}$ & $5.1 \times 10^{4}$
\\
$0.01$ & $0.1834$ & $1.3 \times 10^{2}$ & $3.0 \times 10^{2}$
\\
$0.01$ & $0.1844$ & $8.1 \times 10^{1}$ & $1.1 \times 10^{2}$
\\
\hline
$0.05$ & $0.1993$ & $1.7 \times 10^{2}$ & $-1.7 \times 10^{3}$
\\
$0.05$ & $0.1996$ & $1.6 \times 10^{2}$ & $-2.7 \times 10^{3}$
\\
$0.05$ & $0.1998$ & $1.6 \times 10^{2}$ & $-8.1 \times 10^{3}$
\\
$0.05$ & $0.1999$ & $1.6 \times 10^{2}$ & $3.1 \times 10^{5}$
\\
\hline
$0.10$ & $0.2180$ & $1.6 \times 10^{2}$ & $-6.6 \times 10^{2}$
\\
$0.10$ & $0.2184$ & $1.6 \times 10^{2}$ & $-2.6 \times 10^{3}$
\\
$0.10$ & $0.2185$ & $1.5 \times 10^{2}$ & $-8.1 \times 10^{3}$
\\
$0.10$ & $0.2186$ & $1.5 \times 10^{2}$ & $8.0 \times 10^{3}$
\\
\hline
$0.15$ & $0.2348$ & $2.0 \times 10^{2}$ & $-4.8 \times 10^{2}$
\\
$0.15$ & $0.2352$ & $2.0 \times 10^{2}$ & $-1.7 \times 10^{3}$
\\
$0.15$ & $0.2353$ & $1.9 \times 10^{2}$ & $-3.4 \times 10^{4}$
\\
$0.15$ & $0.2354$ & $1.9 \times 10^{2}$ & $5.9 \times 10^{3}$
\\
\hline
$0.20$ & $0.2494$ & $8.1 \times 10^{2}$ & $-1.8 \times 10^{3}$
\\
$0.20$ & $0.2495$ & $8.5 \times 10^{2}$ & $-2.4 \times 10^{4}$
\\
$0.20$ & $0.2496$ & $7.7 \times 10^{2}$ & $2.3 \times 10^{3}$
\\
\hline
\hline
\end{tabular}
\label{tab:timescale}
\end{center}
\end{table}

These two timescales can also be derived analytically in Newtonian
gravity.  The azimuthal component of the Newtonian Navier-Stokes
equation is
\begin{equation}
\frac{\partial \Omega}{\partial t} = 
  \frac{\nu}{\varpi^{3}} \frac{\partial}{\partial \varpi} 
  \left( \varpi^{3} \frac{\partial \Omega}{\partial \varpi} \right),
\end{equation}
where $\nu$ is a shear viscosity.  We assume the time dependence of
the angular velocity as $\Omega(t) = \Omega_{\rm eq} e^{- i \sigma t}$
at the derivation from the equilibrium due to viscosity, the timescale
to change the angular velocity profile to the uniform rotation is 
\begin{equation}
\tau_{\rm ang} = \sigma^{-1}= \frac{(\varpi^2 + d^2)^2}{8 d^2 \nu} 
  >  \frac{d^2}{8 \nu} 
  = \frac{R^2}{8 \nu} \frac{\Omega_c}{\Omega_c - \Omega_s}
,
\end{equation}
where $\Omega_{\rm s}$ is the equatorial surface angular velocity.
Note that we adopt the $j$-constant rotational profile to the angular 
velocity (Eq.~[\ref{eqn:RLaw_Newton}] in Newtonian gravity).  The
$e$-folding time of a compressible uniformly rotating star (Eq.~[A20]
of Ref.~\citep{LS95}) based on the Navier-Stokes equation is
\begin{equation}
\tau_{\rm bar} = 
  \frac{\kappa_n R^2}{5 \nu}
  \frac{\beta_{\rm sec}}{\beta - \beta_{\rm sec}},
\end{equation}
where $\kappa_n$ is a structure constant depends on the polytropic
index (Table~1 of Ref. \citep{LRS94}; $\kappa_n = 0.65345$ for $\Gamma
= 2$), $\beta_{\rm sec}$ a critical value of $\beta$ for the secular
bar mode instability, $R$ is the equatorial radius of the star.

Based on the analytical estimation of the two timescales, the
timescales used in our numerical results should be described as
\begin{eqnarray}
\tau_{\rm ang} &\approx& \varepsilon_{\rm org}^{-1}
,
\\
\tau_{\rm bar} &\approx& \varepsilon_{\rm org}^{-1} 
  \left( \frac{\Omega_c - \Omega_s}{\Omega_c} \right)
  \left( \frac{\beta_{\rm sec}}{\beta - \beta_{\rm sec}} \right).
\label{eqn:timescale_bar}
\end{eqnarray}
We confirm that differential rotation stabilizes the star from
viscosity driven instability.  Also Eq.~(\ref{eqn:timescale_bar})
shows that the timescale of the bar becomes short when the rotational
profile varies due to viscosity.  The critical value of $T/W$ changes
from the one computed in a fixed rotation profile, but the deviational
ratio of $T/W$ is roughly the same order as the one of rotational
profile, which means $\approx \varepsilon_{\rm org}$.

\section{Conclusions}
\label{sec:Discussion}
We have studied the viscosity driven instability in both uniformly and 
differentially rotating polytropic star by means of iterative evolution
approach in general relativity. We have focussed on the determination
of the critical value of viscosity driven instability.

We find that relativistic gravitation stabilizes the star from the
viscosity driven instability, with respect to Newtonian
gravitation.  Also the critical value is not sensitive to the
stiffness of polytropic equation of state for a given compactness of
the star.  In a previous study devoted to compressible stars,
\citet{BFG98} showed that relativistic gravitation does stabilize the
uniformly rotating polytropic stars by investigating the mass-shedding
sequence.  Our study shows the concrete value of $(T/W)_{\rm crt}$ in
the case of a uniformly rotating star. Also we have improved the
numerical code with respect to the study \citep{BFG98} by making use
of surface-fitted coordinates.

Besides we have found that differential rotation also stabilizes the
star from the viscosity driven instability.  If we fix the compactness
of the star, we find a significant increase of the critical value of
$T/W$, which supports the above statement.  We also confirm this
statement by changing the rotation profile due to viscosity and find
that the differential rotation still stabilizes the threshold of viscosity
driven instability.  However to confirm the statement in
differentially rotating stars, the study of other approaches such as
implicit hydrodynamical evolution or eigenmode analysis should be
necessary and helpful.

Finally let us mention the characteristic amplitude and frequency of
gravitational waves emitted throughout the viscosity driven secular
bar mode instability, which produces quasi-periodic gravitational
waves detectable in ground based interferometers.  The characteristic
amplitude $h$ of gravitational waves estimated from the evolution of a
Jacobi-ellipsoid to a Maclaurin spheroid is (Eq.~[4.2] of
Ref.~\citep{LS95})
\begin{widetext}
\begin{equation}
h \approx 9.1 \times 10^{-21} \left( \frac{30 {\rm Mpc}}{d} \right) 
\left( \frac{M}{1.4 M_{\odot}} \right)^{3/4} 
\left( \frac{R}{10 {\rm km}} \right)^{1/4} f^{-1/5}, 
\end{equation}
\end{widetext}
where $d$ is the distance to the source and the characteristic
frequency $f = \Omega / \pi$ is $f \gtrsim 1000 {\rm [Hz]}$, depending
on $T/W$ of the star.  Note that the frequency increases throughout
this process.  Although the frequency regime of the source is slightly
higher than the best sensitivity regime of the ground based detectors
to follow all the deformation process, we may have a chance to detect
them when it happens in the Virgo cluster, for example.

\acknowledgments
We would like to thank Silvano Bonazzola, Philippe Grandcl\'ement and
J\'er\^ome Novak for their valuable advice.  This work was supported
in part by the PPARC grant (No.~PPA/G/S/2002/00531) at the University of
Southampton, by the Observatoire de Paris, and by the WG3:
Gravitational waves and related studies: Future activities, European
Network of Theoretical Astroparticle Physics.  Numerical computations
were performed on the bi-Xeon machines in the Laboratoire de l'Univers
et de ses Th\'eories, Observatoire de Paris and in the General
Relativity Group at the University of Southampton.



\begin{thebibliography}{99}
\bibitem{footnote}
In this paragraph, we restore the gravitational constant for
convenience of discussion.
\end{thebibliography}
\end{document}